\documentclass[12pt,onecolumn]{extarticle}
\usepackage{newlfont}
\usepackage{amsmath}
\usepackage{graphics}
\usepackage{float}
\usepackage{graphicx}
\usepackage{epsfig}
\usepackage{multirow}
\usepackage{multicol}
\usepackage{verbatim}
\usepackage{pdflscape}
\usepackage{cite}
\usepackage{array}
\usepackage{lineno}
\newcolumntype{H}{>{\setbox0=\hbox\bgroup}c<{\egroup}@{}}
\usepackage{setspace}
\title{Mass Spectrum, Root Mean Square Radii, Form Factors and Charge Radii of Mesons}
\author{Nosheen Akbar, Bushra Shafaq$^\ddag$, Sarwat Zahra$^\S$, Amna Mir$^\ast$ \\
\textit{$\ast$Department of Physics, COMSATS University Islamabad, Lahore Campus,}\\
{Lahore, Pakistan.}\\
\textit{$\ddag$CHEP University of the Punjab, Lahore, Pakistan}\\
\textit{$\S$ University of Education, Lahore, Pakistan}\\
\textit{corresponding author e mail: nosheenakbar@cuilahore.edu.pk}}
\date{}

\begin{document}
\maketitle
\begin{abstract}
In this work, properties of charmonium, bottomonium and charmed-bottom mesons for S, P, D states are calculated with the selection of suitable trial wave functions for the non-relativistic potential model along with the incorporation of spin interactions. These trial wave functions are tested by calculating the masses of ground, radial and orbital excited states of mesons with the variational method. The mass predictions show good agreement with the available experimental data and theoretical predictions found by different methods. This indicates that these trial wave functions are appropriate for further study on mesons. These trial wave functions are used to calculate RMS radii, form factors and charged radii.  Results show that RMS radii and charged radii have inverse relation with the masses of mesons, i.e., heavier mesons have smaller radii and vice versa. Momentum dependence of the form factors is studied graphically.
\end{abstract}
%%% ----------------------------------------------------------------------
%%% ----------------------------------------------------------------------
% \textbf{PACS numbers}:  12.39.Pn, 12.38.-t, 12.39.jh \\
\maketitle
%\linenumbers
%\doublespacing
\section*{I. Introduction}

Wave functions (WFs) and phenomenological potential models play very important role in the investigation of hadrons in QCD (Quantum Chromodynamics) \cite{Brodsky}. Mostly used phenomenological potential model for the study of mesons in the non-relativistic limit is the linear plus coulombic potential model.  This non-relativistic potential model is used in Refs. \cite{00601-015,TAPASHI} to find solutions ( or wave functions)  of Schrodinger equation for the study of mesons. It is necessary to find appropriate wave functions for the measurements of static as well as dynamic properties of mesons. With help of these wave functions mass spectrum, radii, decay constants, form factors, etc can be calculated.

Different methods are used to find wave functions such as perturbation theory \cite{Fernandez,Ghatak}, variationally improved perturbation theory \cite{Aitchison,You,Fernandez2003}, the WKB method \cite{Ghatak}, Dalgarno method \cite{TAPASHI}, and the variational method \cite{Ghatak,Raghav}. In Ref.\cite{00601-015}, variational method is used by choosing the trial WFs as product of Laguerre polynomial, exponential function and inter-quark distance. Then these wave functions are tested by calculating the masses and decay constants of mesons for 1S and 1P states. The results prove the success of variational method. With these wave functions, form factors can be measured that play important role in giving information about the structure of hadrons and depends on the momentum transform of WFs. Theoretically, form factors of mesons are calculated with two loop potential model \cite{sharma}, Variationally improved perturbation theory \cite{0069-0085}, contact interaction \cite{wang,2301.11881}, Schwinger-Dyson equations (SDEs)\cite{Chang,Maris,Bhagwat} within the functional approach, and lattice QCD \cite{Alexandrou,Davies,Shultz,Karthik}; while experimental work on electromagnetic FFs of mesons is presented in Ref. \cite{Horn,Amendolia}. Charged radii of mesons using the form factors is calculated in \cite{sharma,0069-0085,2301.11881,wang}.

In present work, a simpler approach is used to find the form factors and charge radii. In this work, variational method is applied on the trial WFs (chosen in little bit simple forms) for 1S, 1P, 1D, 2S, 2P, 2D states of mesons to calculate the masses and radii of charmonium, bottomonium and $B_c$ mesons. For this, non-relativistic potential model is considered along with the addition of spin interaction term. The agreement of the calculated masses with the experimental and theoretical available data shows that selection of these trial WFs is appropriate. Using these WFs, form factors (FFs) in a range of momentum square $0 \leq Q^2 \leq 2 \textrm{GeV}^2$ and charged radii ($r_c$) are calculated for ground and radially excited $S$, $P$, and $D$ states of mesons with spin = 0, 1.

The paper is organized in four sections. Section 1 is introductory section. In the section 2 of this paper, trial WFs for the mesons in the ground and excited states are defined. These trial WFs are used to find the masses and radii of mesons by variational method with non-relativistic potential model. Methodology for the calculation of FFs and $r_c$ is discussed in section 3. Results are reported in Section 4 and discussed in section 5 along with the concluding remarks.
%\begin{singlespace}
\section*{2. Wave Functions of Mesons for Non-Relativistic Potential Model}
%\end{singlespace}
Properties of mesons can be found with help of Radial Schr$\textrm{\"{o}}$dinger equation defined as
\begin{equation}
 H \psi (r, \alpha) = E \psi (r, \alpha)
\end{equation}
For mesons, the Hamiltonian ($H$) can be written as
\begin{equation}
H = \frac{-\hbar^2}{2\mu}\nabla^2 - V(r) - \frac{L (L + 1)}{2 \mu r^{2}}.
\end{equation}
Here, V(r) is used for potential of mesons. In Ref. \cite{barnes05}, potential of mesons is modelled as:
\begin{equation}
V(r)= \frac{-4\alpha _{s}}{3r}+ \beta r + \frac{32\pi \alpha_s}{9 m_q m_{\overline{q}}} (\frac{\sigma}{\sqrt{\pi}})^3 e^{-\sigma ^{2}r^{2}} \textbf{S}_{q}. \textbf{S}_{\overline{q}}.
\end{equation}

In the potential model, the parameter $\alpha_s$ is the strong coupling constant and $\beta$  is the string tension. $\textbf{S}_{q}. \textbf{S}_{\overline{q}}$ is equal to $\frac{S(S+1)}{2}-\frac{3}{4}$ as defined in \cite{barnes05,Nosheen14}. In case of charmonium mesons, $\alpha_s$, $\beta$, $\sigma$ are taken equal to 0.5461, 0.1425 GeV, 1.0946 $\text{GeV}^2$ respectively  \cite{barnes05,Nosheen14}. For bottomonium mesons, $\alpha$, $\beta$, $\sigma$ are taken equal to 0.36, 0.1340 GeV, 1.34 $\text{GeV}^2$ respectively as in Ref.\cite{Nosheen17}. For charmed bottom mesons, $\alpha_s$, $\beta$, $\sigma$ are equal to 0.48, 0.137 GeV, 1.0946 $\text{GeV}^2$ respectively as in Ref. \cite{Nosheen19}. Mass of b-quark and c-quark are taken equal to 4.825 GeV and 1.4794 GeV as used in Refs. \cite{Nosheen14,Nosheen17,Nosheen19}. $E$ is the energy of the meson and depends on the WF. For the radially ground states (i.e. n=1) of $c\overline{c}$, $b\overline{b}$ and $B_c$ mesons,  trial WFs are considered in the following form:
\begin{equation}
 \psi_{1l}(r,a) = \frac{c_{n_1}}{\sqrt{4 \pi}} r^{l+1} e^{-a^2 r^2}
\end{equation}
For the radially excited states (i.e. n = 2), the trial WFs are chosen as:
\begin{equation}
 \psi_{2l}(r,a) = \frac{c_{n_2}}{\sqrt{4 \pi}} r^{(l+1)} (1 - a r) e^{-\frac{(l+1) a^2 r^2}{2}}.
\end{equation}

Here, $c_{n_1}$ and $c_{n_2}$ are the normalization constants. $l$ is orbital quantum number and its value is equal to 0,1,2,... for states $S$, $P$, $D$,...respectively. $a$ is the variational parameter and its value is found by using the following condition:
\begin{equation}
\frac{\partial}{\partial a} (\frac {\left(\langle\psi_{1l}(r, a)\mid H\mid \psi_{1l}(r,a)\rangle\right)}{\langle\psi_{1l}(r, a)\mid \psi_{1l}(r, a)\rangle}) = 0
\end{equation}
With this variational parameter $a$, eigen WFs and eigen energies of mesons are calculated. Adding the mass of quark and anti-quark in the energy eigen values, mass spectrum of mesons is obtained. The normalized eigen WFs are then used to calculate RMS radii and form factors of mesons. RMS radii of mesons are calculated with the following expression as defined in Ref. \cite{2023}
\begin{equation}
\sqrt{\langle r^{2}\rangle} = \sqrt{\int \psi^{\star} r^{2} \psi dr}.\label{P25}
\end{equation}
\section*{3. Form factor (FF) and Charge Radii ($r_c$)}
The FF is defined in Refs.\cite{Stanley} \cite{1608.06896} as:
\begin{equation}
F(Q^2)= \sum_{j=1}^2 (\frac{e_j}{Q_j})\int^{\infty}_0 4 \pi r \mid \psi(r) \mid^2 \sin(Q_j r) dr \label{star}
\end{equation}
where
\begin{equation}
Q_j=\sum_{j \neq i}m_i Q/\sum_{j=1}^2 m_j,
\end{equation}
$e_j$ is the charge on quark and antiquark. $\sin(Q_j r)$ can be expanded in series form as
\begin{equation}
\sin(Q_j r)=(Q_j r)-\frac{(Q_j r)^3}{3 !}+\frac{(Q_j r)^5}{5 !}-\frac{(Q_j r)^7}{7 !}+----
\end{equation}
With wave functions defined in eqs. (4,5), $F(Q^2)$ are calculated for 1S, 1P, 1D,  2S, 2P, 2D states of $c\overline{c}$, $b\overline{b}$ and $B_c$ mesons. In case of neutral charge mesons( $c\overline{c}$ and $b\overline{b}$), calculations give zero value for $F(Q^2)$ because the form factor due to quark cancel the form factor due to anti-quark. To avoid this zero value of $F(Q^2)$, the charge of anti-quark is considered equal to 0. The similar methodology is applied in Refs. \cite{2301.11881} \cite{1809.06475}. Graphs of normalized $F(Q^2)$ versus $Q^2$ are shown in Figs.(1,2).

$r^2_c$ for mesons can be calculated by using the following expression as defined in \cite{1608.06896, 0519-0534}
\begin{equation}
\langle r_c^2 \rangle = -6 \frac{d}{d Q^2}\mid F(Q^2) \mid_{Q^2=0}
\end{equation}
\section*{4. RESULTS}
Masses of 1S, 1P, 1D, 2S, 2P, 2D states of charmonium, bottomonium,and charmed-bottom mesons are reported in Tables (1-3 )with spin-0 and spin-1. RMS radii and charge radii are reported in Tables (4-6). In Figs. (1,2), the variation in FFs with $Q^2$ is shown graphically.
%\clearpage
\begin{table}[h]
\begin{center}
\caption{\label{demo-table}Masses of charmonium mesons in GeV. Our calculated results are rounded to 4 decimal places.}
\begin{tabular}{|l|l|l|l|l|l|l|l|}
 \hline
& Variational & \multicolumn{6}{c|}{Mass}  \\
 \cline{3-8}  & parameter & This work & Experimental \cite{pdg} & \cite{2301.12388} & \cite{2011}   & \cite{2018}& \cite{variational} \\

 & & & & & & & \\ \hline
\multirow{1}{*}{(1$^{1}S) \eta_{c}$}& 0.4860 & 3.0330 &2.9839$\pm0.0004$ & 2.9644 & 2.981 &3.068 & 2.980\\
\hline
\multirow{1}{*}{(1$^{3}S)J/\psi$}& 0.4207 & 3.117 &3.0969$\pm0.000006$& 3.0964 &3.096 & - &0.096\\
\hline
\multirow{1}{*}{(1$^{1}P)h_{c}$}& 0.3296 & 3.5260 &3.52538$\pm0.00011$& 3.4161 & 3.525 &3.534&-\\
\hline
\multirow{1}{*}{(1$^{3}P)\chi$}& 0.3246 & 3.5319 &3.55617$\pm0.00007$& 3.4358 &3.555 & - & -\\
\hline
\multirow{1}{*}{(1$^{1}D)\eta_{c2}$}& 0.2919 & 3.8040 & - & 3.6751 & 3.807 &3.802& -\\
\hline
\multirow{1}{*}{(1$^{3}D)\psi$}& 0.2915 & 3.8044 &3.77313$\pm0.0004$& 3.6881 & 3.783& -& -\\
\hline
\multirow{1}{*}{(2$^{1}S) \eta_{c}$}& 0.40908 & 3.6236 &3.6375$\pm$0.0011&3.5078 &3.635 &3.638 &3.624 \\ \hline
\multirow{1}{*}{(2$^{3}S) J/\psi$}& 0.3940 & 3.6678&3.68610$\pm0.00006$&  3.605 &3.685 & -& 3.727\\
\hline
\multirow{1}{*}{(2$^{1}P) h_{c}$}& \multirow{1}{*}0.2610& 3.9335 & - & 3.8774 & 3.926 &3.936& -\\  \hline
\multirow{1}{*}{(2$^{3}P)\chi$}& \multirow{1}{*}0.2578 & 3.9413 &- & 3.9011& 3.949 &-& - \\
\hline
\multirow{1}{*}{(2$^{1}D)\eta_{c2}$}& \multirow{1}{*}0.2024& 4.1573 & -& - &4.196 & 4.150 & - \\
\hline\multirow{1}{*}{(2$^{3}D)\psi$}& \multirow{1}{*}0.2020& 4.1582 & - & - &4.150 & -& -\\
\hline
\end{tabular}
\end{center}
\end{table}

\begin{table}[h!]
\begin{center}
\tabcolsep=6pt \fontsize{10}{12}\selectfont
\caption{\label{demo-table}Masses of bottomonium mesonsin GeV. Our calculated results are rounded to 4 decimal places.}
\begin{tabular}{|l|l|l|l|l|l|l|l|l|l|}
 \hline
 state & Variational & \multicolumn{7}{c|}{Mass} \\
  & Parameter & This work & Experimental\cite{pdg} &  \cite{2301.12388} & \cite{2011} &\cite{2018}& \cite{Nosheen17} & \cite{variational}\\
\hline
\multirow{1}{*}{(1$^{1}S)\eta_{b}$}& \multirow{1}{*}0.7828&  9.5535 &9.3987$\pm$0.002 & 9.5615 &9.398&9.398 &9.5079 & 9.452\\
\hline
\multirow{1}{*}{(1$^{3}S)\Upsilon_{b}$}& \multirow{1}{*}0.7571 & 9.5722&9.4603$\pm0.00026$& 9.6478 &9.478 & 9.460 &9.5229 & 9.480\\
\hline
\multirow{1}{*}{(1$^{1}P)h_{b}$}& \multirow{1}{*}0.5129& 9.9373 &9.8993$\pm0.0008$& 9.9324 &9.900 &9.894&9.9279 & -\\  \hline
\multirow{1}{*}{{(1$^{3}P)\chi$}}& \multirow{1}{*}0.5096 & 9.9391 & \vtop{\hbox{\strut 9.91221}\hbox{\strut $\pm0.00026$}\hbox{\strut $\pm0.00031$}}& 9.9389 & 9.912&9.858& 9.9232 &-\\ \hline
\multirow{1}{*}{{(1$^{1}D)\eta_{b2}$}}& \multirow{1}{*}0.4425& 10.1398 & - &- & 10.163 & -&10.1355 & -\\ \hline
\multirow{1}{*}{(1$^{3}D)\Upsilon$}& \multirow{1}{*}0.4422& 10.1399 &-&-& 10.161 && 10.1548 & -\\ \hline
\multirow{1}{*}{(2$^{1}S)\eta_{b}$}& \multirow{1}{*}0.62615&  9.9980  & - & - & 9.990 & 10.017 & 10.0041 & 10.030\\ \hline
\multirow{1}{*}{(2$^{3}S)\Upsilon$}& \multirow{1}{*}0.6215& 10.0052 &10.02326$\pm$0.00031& 10.0167 & 10.023 &10.356&10.0101 & 10.055\\ \hline
\multirow{1}{*}{(2$^{1}P)h_{b}$}& \multirow{1}{*}0.3924&10.2210 &10.2598$\pm0.0012$&10.2161 & 10.260 &10.259& - & -\\ \hline
\multirow{1}{*}{(2 $^{3}P)\chi$}& \multirow{1}{*}0.3909&10.2288 &\vtop{\hbox{\strut 10.26865}\hbox{\strut $\pm0.0002$}\hbox{\strut $\pm0.0005$}}& - &  10.2232 &10.255& - & -\\ \hline
\multirow{1}{*}{(2 $^{2}D)\eta_{b2}$}& \multirow{1}{*}0.301959&10.3780 & & - & - & 10.450& - & -\\ \hline
\multirow{1}{*}{(2 $^{3}D)\Upsilon$}& \multirow{1}{*}0.30170&10.3783 & - & - & 10.443 &10.442& - & -\\ \hline
\end{tabular}
\end{center}
\end{table}

\clearpage

\begin{table}[h!]
\begin{center}
\caption{\label{demo-table} Masses of $B_c$ mesons in GeV. Our calculated results are rounded to 4 decimal places.}
\begin{tabular}{|l|l|l|l|l|l|l|l|l|l|}
 \hline
state & Variational & \multicolumn{5}{c|}{Mass}\\
 & parameter & This work & Experimental \cite{pdg} & \cite{2011} & \cite{2020} & \cite{Nosheen19} \\
\hline
\multirow{1}{*}{(1$^{1}S) B_{c}$}& \multirow{1}{*}0.5548& 6.3164 &6.27447$\pm0.00027$$\pm0.00017$ &6.272&6.277 & 6.274 \\
\hline
\multirow{1}{*}{(1$^{3}S) B_{c}$}& \multirow{1}{*}0.5241 & 6.3494& - & 6.333 & - &6.314\\
\hline
\multirow{1}{*}{(1$^{1}P) B_{c}$}& \multirow{1}{*}0.3837& 6.7458 & - & 6.743&6.689&6.744\\  \hline
\multirow{1}{*}{(1$^{3}P) B_{c}$}& \multirow{1}{*}0.38044 & 6.7488 & -& 6.761 & - &6.753 \\
\hline
\multirow{1}{*}{(1$^{1}D) B_{c}$}& \multirow{1}{*}0.3373& 6.9907 &- & 7.025 &6.753& 6.984\\
\hline
\multirow{1}{*}{(1$^{3}D) B_{c}$}& \multirow{1}{*}0.3370& 6.9910 & -& 7.021 & - & 6.964\\
\hline
\multirow{1}{*}{(2$^{1}S) B_{c}$}& \multirow{1}{*}0.4699& 6.8353 &6.8712$\pm0.001$& 6.882&6.868&6.841\\
\hline
\multirow{1}{*}{(2$^{3}S) B_{c}$}& \multirow{1}{*}0.4637& 6.8504 &-& 7.226&7.109&6.855\\
\hline
\multirow{1}{*}{(2$^{1}P) B_{c}$}& \multirow{1}{*}0.3000& 7.1000 &-& 7.134 & - &7.098\\
\hline
\multirow{1}{*}{(2$^{3}P) B_{c}$}& \multirow{1}{*}0.2982& 7.1034 &-&7.157& - &7.111\\
\hline
\multirow{1}{*}{(2$^{1}D) B_{c}$}& \multirow{1}{*}0.2325& 7.2940 &- & 7.399&6.742 & 7.293\\
\hline\multirow{1}{*}{(2$^{3}D) B_{c}$}& \multirow{1}{*}0.2322& 7.2945 &-&7.392& - &7.280\\
\hline
\end{tabular}
\end{center}
\end{table}

\begin{table}[h!]
\begin{center}
\caption{\label{demo-table}RMS radii and charge radii of charmonium mesons in $fremi$.}
\begin{tabular}{|l|l|l|l|l|l|l|l|}
 \hline
state & \multicolumn{3}{c|}{$ \sqrt{r^{2}}$ } & \multicolumn{4}{c|}{Charge Radii} \\
 & This work & \cite{Nosheen14} & \cite{soni} & This work &\cite{2301.11881} & \cite{1809.06475} & \cite{LC209}\\ \hline

(1$^{1}S)\eta_{c}$ & 0.3515& 0.3655 &0.463 & 0.2269 & 0.20 & 0.207 & 0.249\\ \hline
(1$^{3}S)J/\psi$ & 0.40606 & 0.4143 &- & 0.2621 & -& 212 & 0.257 \\ \hline
(1$^{1}P)h_{c}$ & 0.6690 & 0.6738 &0.706 & 0.3958 &-&0.265 & 0.322\\ \hline
(1$^{3}P)\chi$ & 0.6793 &0.7173 & - & 0.4019 & 0.43 &- &\\ \hline
(1$^{1}D)\eta_{c2}$ & 0.8939 & 0.8984& 0.902 & 0.5068 & - & - & -\\ \hline
(1$^{3}D)\psi$ & 0.8951 & 0.9179 & - & 0.5075 &- &- & -\\ \hline

(2$^{1}S)\eta_{c}$& 0.8390 &0.8328 & 0.887 & 0.5130 &- &0.386 & 0.381\\ \hline
(2$^{3}S)J/\psi$& 0.8711 & 0.8627 & - & 0.5326 & -&0.387 & 0.385\\ \hline
(2$^{1}P)h_{c}$& 1.07181 & 1.0697& 1.080 & 0.6494 &- &- & -\\  \hline
(2$^{3}P)\chi$ & 1.0850 & 1.1107& - &0.6574 & -& -& -\\ \hline
(2$^{1}D)\eta_{c2}$& 1.2603 & 1.2595& 1.246 & 0.7511 & -& -& -\\ \hline
(2$^{3}D)\psi$ & 1.2627 & 1.2788 & - & 0.7525 &- &- & -\\ \hline
\end{tabular}
\end{center}
\end{table}

\clearpage

\begin{table}[h!]
\begin{center}
\caption{\label{demo-table}RMS radii and charge radii of Bottomonium mesons in $fermi$.}
\begin{tabular}{|l|l|l|l|l|l|l|}  \hline

state & \multicolumn{3}{c|}{$\sqrt{r^{2}}$} &  \multicolumn{3}{c|}{Charge Radii}\\
state & This work &\cite{soni} & \cite{Nosheen17} & This work &\cite{2301.11881} &\cite{LC209} \\ \hline

$\eta_{b}(1^{1}S)$ & 0.2182 & 0.259&0.2265 & 0.1408 &0.07 & 0.152 \\ \hline
$\Upsilon_{b}(1 ^{3}S)$& 0.2256 & &0.2328 & 0.1456& - & 0.154\\ \hline
$h_{b}(1 ^{1}P)$& 0.430 & 0.407& 0.4347 & 0.2544& - & 0.196 \\ \hline
$\chi(1^{3}P)$& 0.4327 & - & 0.4375 & 0.2560& 0.39 & - \\ \hline
$\eta_{b2}(1^{1}D)$& 0.5897 &0.525& 0.5933 & 0.3343 & - & -\\ \hline
$\Upsilon(1^{3}D)$&0.5901& - & 0.5942 & 0.3346& - & -\\ \hline

$\eta_{b}(2^{1}S)$& 0.5481 & 0.512 & 0.5408 & 0.3351 & - & 0.232\\ \hline
$\Upsilon_{b}(2^{3}S)$ & 0.5522 & - & 0.5448 & 0.3376 & - & 0.233\\ \hline
$h(2^{1}P)$ & 0.7128 &0.629 & 0.7114 & 0.4318 & - & -\\ \hline
$\chi(2^{3}P)$ & 0.7155 & - &0.7139 & 0.4335 & - & -\\ \hline
$\eta_{b2}(2^{2}D)$ & 0.8447 & 0.729 & 0.8447 & 0.5034  & - & -\\ \hline
$\Upsilon(2^{3}D)$ & 0.8454 & - & 0.8457 & 0.5039 & - & -\\ \hline

\end{tabular}
\end{center}
\end{table}

\begin{table}[h!]
\begin{center}
\caption{\label{demo-table}RMS radii and charge radii of charmed-bottom mesons. radii are rounded to 4 decimal places}
\begin{tabular}{|l|l|l|l|l|l|l|}
 \hline
state & \multicolumn{3}{c|}{$\sqrt{r^{2}}$} & \multicolumn{3}{c|}{Other theoretical work}\\
 & This work & \cite{soni} & \cite{Nosheen19}& This work &\cite{2301.11881} &\cite{LC209} \\ \hline

$B_c(1^{1}S_0 )$ & 0.3079 & 0.373 & 0.318 & 0.1914 & 0.17 & 0.282\\ \hline
$ B_c(1^{3}S_1)$ & 0.3259 & - & 0.334 & 0.2026 & - & 0.286  \\ \hline
$B_c(1^{1}P)$ & 0.5746 & 0.572 & - & 0.3275& - & - \\  \hline
$B_c(1^{3}P)$ & 0.5796 & - & 0.594 & 0.3303  & 0.40 & - \\ \hline
$B_c(1^{1}D )$ & 0.7735 & 0.733 & - & 0.4224& - & - \\ \hline
$B_c(1^{3}D )$ & 0.7742 & - &0.793 &0.4228 & - & - \\ \hline

$B_c(2^{1}S )$ & 0.7304 &0.719 & 0.723 &0.4302 & - & 0.430  \\ \hline
$B_c(2^{3}S)$ & 0.7402 & 0.877 & 0.732 & 0.4359 & - & 0.433 \\ \hline
$B_c(2^{1}P)$ & 0.9324 &0.877 &- &0.5441 &- & 0.364  \\ \hline
$B_c(2^{3}P)$ & 0.9379 & - &0.940 &0.5474 &- & - \\ \hline
$B_c(2^{1}D)$ & 1.0972 & 1.013 & - &0.6299 & & - \\ \hline
$B_c(2^{3}D)$ & 1.0985 & - &1.107 &0.6306 &- & - \\ \hline

\end{tabular}
\end{center}
\end{table}

\clearpage

\begin{figure}[htbp]
\centering
%\begin{center}
\begin{tabular}{ccc}
\includegraphics[width=0.6\textwidth]{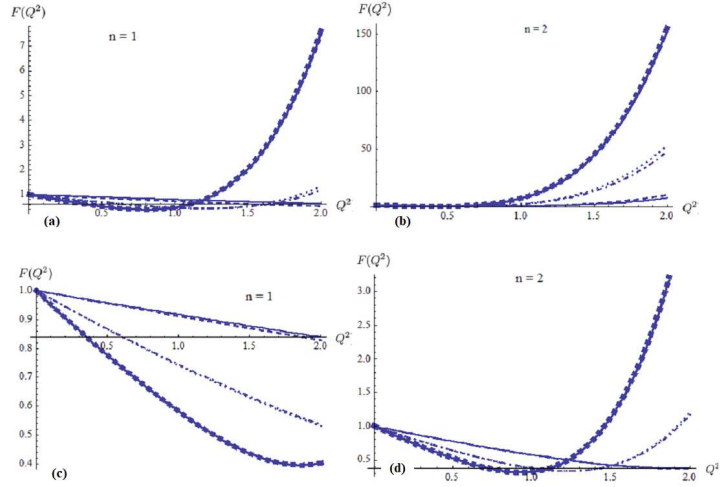}
\end{tabular}
\caption{(a, b). Form factors of Charmonium (c, d). Form factors of bottomonium mesons versus $Q^2$. Form factors of $^1 S_0$ state are represented by solid line curve, $^3 S_1$ state form factors are represented by dashed line curve, $^1 P_1$ state form factors are represented by dots-dashed line curve, $^3 P$ state form factors are represented by dotted line curve, $^1 D_2$ state form factors are represented by thick line curve, $^3 D$ state form factors are represented by squared line curve.}
%\end{center}
\end{figure}

%\begin{figure}
%\begin{center}
%\epsfig{file=charmonium.eps,width=0.8\linewidth,clip=}\caption{Form factors of Charmonium mesons versus $Q^2$. Form factors of $^1 S_0$ state are represented by solid line curve, $^3 S_1$ state form factors are represented by dashed line curve, $^1 P_1$ state form factors are represented by dots-dashed line curve, $^3 P$ state form factors are represented by dotted line curve, $^1 D_2$ state form factors are represented by thick line curve, $^3 D$ state form factors are represented by squared line curve.}
%\end{center}
%\end{figure}

%\begin{figure}
%\begin{center}
%\epsfig{file=bottomonium.eps,width=0.8\linewidth,clip=}\caption{Form factors of Bottomonium mesons versus $Q^2$. Form factors of $^1 S_0$ state are represented by solid line curve, $^3 S_1$ state form factors are represented by dashed line curve, $^1 P_1$ state form factors are represented by dots-dashed line curve, $^3 P$ state form factors are represented by dotted line curve, $^1 D_2$ state form factors are represented by thick line curve, $^3 D$ state form factors are represented by squared line curve.}
%\end{center}
%\end{figure}

\begin{figure}
\begin{center}
\epsfig{file=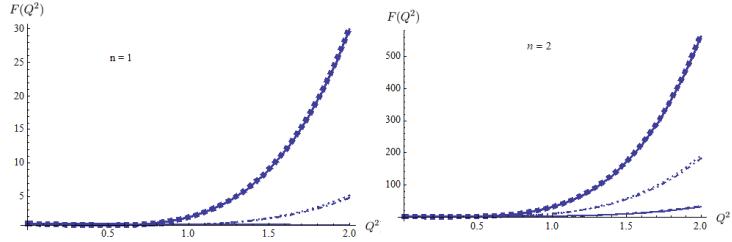,width=0.6\textwidth}
\caption{Form factors of $B_c$ meson versus $Q^2$. Form factors of $^1 S_0$ state are represented by solid line curve, $^3 S_1$ state form factors are represented by dashed line curve, $^1 P_1$ state form factors are represented by dots-dashed line curve, $^3 P$ state form factors are represented by dotted line curve, $^1 D_2$ state form factors are represented by thick line curve, $^3 D$ state form factors are represented by squared line curve.}
\end{center}
\end{figure}

\section*{5. DISCUSSION AND CONCLUSIONS}
Observing the masses of different states of mesons reported in Tables(1-3), we conclude that mass is increasing with orbital and radial excitations. It is observed that our calculated meson's masses are very close to the experimentally reported \cite{pdg} and theoretically calculated masses by different methods \cite{2301.12388,2011,2018,variational,2020,Nosheen17,Nosheen19}. This shows that our selected trial-wave functions (for non-relativistic Schrodinger equation) are suitable to investigate the dynamical properties of mesons like RMS radii, FFs, $r_c$, decay widths, etc.
Calculated RMS radii and charge radii are reported in Tables (4-6).

In Figs. (1,2), the variation in FFs with $Q^2$ is shown graphically. From the graphs, it is observed that the FFs (of mesons) decrease with increasing values of $Q^2$. Similar behaviour for ground state $B_c$, $D$ and $D_s$ mesons was observed in Ref. \cite{0069-0085}. For S = 0, 1 with same value of $l$, the results are very close to each other, so in the Figures these curves overlap each other.

It is observed that the charge radii is increasing with orbital excitations i.e. the radii of $P$ state meson is larger than $S$ state. The same behaviour is observed in Ref.\cite{1809.06475,Dudek}. Similarly, the radii is increasing with radial excitations; i.e. \\
$\langle r^2 (3S)\rangle > \langle r^2( 2S) \rangle > \langle r^2 (1S)\rangle$ \\
$\langle r^2 (3P)\rangle > \langle r^2( 2P) \rangle > \langle r^2 (1P)\rangle$ \\
$\langle r^2 (3D)\rangle > \langle r^2( 2D) \rangle > \langle r^2 (1D\rangle$ \\

It is also observed that the radii is decreasing toward heavier mesons, i.e., radii ($b\overline{b}$) $<$ radii ($B_c$) $<$ radii ($c\overline{c}$). From this observation, we can conclude that heavier mesons are more close to point particles than the lighter mesons. Similar relation between mass and radii is observed in \cite{1711.00383}.

\end{document}